\documentclass[twocolumn,showpacs,amsmath,amssymb,preprintnumbers,aps,prd,reprint]{revtex4-1}

\usepackage{graphicx}
\usepackage{epsfig}
\usepackage{dcolumn}
\usepackage{bm}
\usepackage{tabularx}
\usepackage{subfigure}
\usepackage{siunitx}
\newcommand{\ord}{\mathcal{O}}

\newcommand{\tn}{\textnormal}

\newcommand{\y}{\gamma}

\newcommand{\lcdm}{$\Lambda$CDM} % For a space following this command use "\ "

\newcommand{\gagg}{$g_{a\y\y}$}

\newcommand{\bei}{\begin{itemize}}
\newcommand{\eei}{\end{itemize}}
\newcommand{\beq}{\begin{equation}}
\newcommand{\eeq}{\end{equation}}
\newcommand{\beqn}{\begin{eqnarray}}
\newcommand{\eeqn}{\end{eqnarray}}
\newcommand{\beqns}{\begin{eqnarray*}}
\newcommand{\eeqns}{\end{eqnarray*}}

\newcommand{\jprlBase}       {Phys.\ Rev.\ Lett.}

\newcommand{\jprl}      [1]  {\jprlBase\ {\bf #1}}
\begin{document}

\title{Modulation Sensitive Search for Non-Virialized Dark-Matter Axions}

\author{J. Hoskins}
\author{N. Crisosto}
\author{J. Gleason}
\author{P. Sikivie}
\author{I. Stern}
\author{N. S. Sullivan}
\author{D. B. Tanner}
\affiliation{University of Florida, Gainesville, FL 32611, USA}

\author{C. Boutan}
\author{M. Hotz}
\author{R. Khatiwada}
\author{D. Lyapustin}
\author{A. Malagon}
\author{R. Ottens}
\author{L. J Rosenberg}
\author{G. Rybka}
\author{J. Sloan}
\author{A. Wagner}\altaffiliation[Currently at ]{Raytheon BBN Technologies, Quantum Information Division, 10 Moulton Street, Cambridge, MA 02138}
\author{D. Will}
\affiliation{University of Washington, Seattle, WA 98195, USA}

%\author{S. Chatopadhyay}
%\author{A. S. Chou}
%\author{A. Sonnenschein}
%\author{W. Wester}
%\affiliation{Fermi National Accelerator Laboratory, Batavia IL 60510, USA}

\author{G. Carosi}
\author{D. Carter}
\affiliation{Lawrence Livermore National Laboratory, Livermore, CA 94550, USA}

\author{L. D. Duffy}
\affiliation{Los Alamos National Laboratory, Los Alamos, NM 87544, USA}

\author{R. Bradley}
\affiliation{National Radio Astronomy Observatory, Charlottesville, VA 22903, USA}

\author{J. Clarke}
\author{S. O'Kelley}
\author{K. van Bibber}
\affiliation{University of California, Berkeley, CA 94720, USA}

\author{E. J. Daw}
\affiliation{University of Sheffield, Sheffield UK}
\date{\today}

\begin{abstract}Non-virialized dark-matter axions may be present in the Milky Way halo in the form of low-velocity-dispersion flows. The Axion Dark Matter eXperiment performed a search for the conversion of these axions into microwave photons using a resonant cavity immersed in a strong, static magnetic field. The spread of photon energy in these measurements was measured at spectral resolutions of the order of 1 Hz and below. If the energy variation were this small, the frequency modulation of any real axion signal due to the orbital and rotational motion of the Earth would become non-negligible. Conservative estimates of the expected signal modulation were made and used as a guide for the search procedure. The photon frequencies covered by this search are 812--852 and 858--892 MHz, which correspond to an axion mass of 3.36--3.52 and 3.55--3.69 $\mu$eV. No axion signal was found, and limits were placed on the maximum local density of non-virialized axions of these masses.
\end{abstract}

\maketitle

%-------------------------------------------------------------------------------------------------------%

\section{\label{sec:intro}INTRODUCTION}%

Due to being long-lived, non-relativistic, and effectively collisionless\cite{Abbot.83,Preskill.83,Dine.83,Ipser.83}, the axion, which stems from the Peccei-Quinn solution to the strong CP problem\cite{Peccei.77.1,Peccei.77.2}, is a compelling candidate for the dark matter component of our universe. Their long lifetime suggests that axions created during the early universe would likely still exist today in detectable quantities, while their low speeds and feeble coupling strengths would be consistent with the $\Lambda$ Cold Dark Matter (\lcdm) model of our universe. Though the axion has extraordinarily weak couplings to the standard model, its decay into two photons by means of
\begin{equation}
 \label{eq:lagg}
 \mathcal{L}_{a\y\y}=g_\y\frac{\alpha}{\pi}\frac{a(x)}{f_a}\textbf{E}\cdot\textbf{B},
\end{equation}
would remain a viable avenue for direct detection. Here $\alpha$ is the fine structure constant, $a(x)$ is the axion field, $f_a$ is the Peccei-Quinn symmetry breaking scale, $g_\y$ is a model dependent coupling constant of order 1, $\mathbf{E}$ is the electric field of one decay photon, and $\mathbf{B}$ is the magnetic field of the other. The Kim-Shifman-Vainshtein-Zakharov (KSVZ) model\cite{Kim.79, Shifman.80} predicts $g_\y=-0.97$, while the Dine-Fischler-Srednicki-Zhitnitskii (DFSZ) model\cite{Dine.81, Zhit.80} predicts $g_\y=0.36$. The Axion Dark Matter eXperiment (ADMX) is a direct detection experiment that searches for relic axions via the interaction governed by Eq.~\ref{eq:lagg}. The axion decay would occur inside an axion haloscope, which consists of a microwave resonator immersed in a static magnetic field, and would be measured by a low-noise, high-gain radio receiver\cite{Sikivie.83,Sikivie.85,ADMX.98,ADMX.09}.

There are two parallel data acquisition channels employed by ADMX, each designed to search for a particular population of axions within the Milky Way halo. These populations are defined based on the nature of their velocity distributions. Of particular interest is the dispersion of each distribution as this property directly impacts the spectral width of the signal seen by ADMX. Axions with an average velocity of magnitude $v$ and a velocity dispersion of magnitude $\delta v$ would generate a signal with an rms frequency deviation, $\delta f$, given by
\begin{equation}
 \label{eq:ffrac}
 \frac{\delta f}{f} = \frac{\delta E}{E} \approx \frac{v\delta v}{c^2},
\end{equation}
where $f$ is the central frequency of the signal, $E$ is the average energy of the axions, and $c$ is the speed of light. Because the rms frequency deviation describes fluctuations about the central frequency which can be either positive or negative, the spectral width of the signal is taken to be $\Delta f = 2 \delta f$. Populations of axions with different velocity dispersions would generate signals with correspondingly different widths. Consequently, using channels with differing resolutions greatly improves ADMX's versatility and detection capabilities.

The Medium Resolution (MR) channel searches for virialized axions, as would be found in the isothermal model of our galaxy. These axions are expected to have reached a steady state of motion within the galaxy, and to follow a Maxwell-Boltzmann distribution with a velocity dispersion of $\ord$(10$^{-3}c$). For 3.5 $\mu$eV axions with $v \lesssim 300$ km/s, the spectral width of an axion signal would be $\ord$(1 kHz). Results from the MR channel searches for 250 Hz and 750 Hz bin widths can be found in Ref.~\cite{ADMX.09}.

As a complement to the MR channel, the High Resolution (HR) channel searches for non-virialized axions and is the focus of this paper. Two possible sources of non-virialized axions are axions which have only recently entered the galactic halo and axions which have been pulled out of tidally disrupted subhalos. In both cases the velocity dispersion of non-virialized axions is significantly lower than that of virialized axions \cite{Sikivie.03,Freese.04,Freese.05}. The lower velocity dispersions of these axions would yield signals with proportionally smaller spectral widths. The actual magnitude of the decrease in spectral width is specific to each source of non-virialized axions. A brief description of both late-infall axions and tidal axions is presented below.

Axions that accreted late in the lifetime of the galaxy would not have had sufficient time to thermalize and would exist as discrete, non-virialized streams flowing into and out of the galaxy\cite{Duffy.08}. Flows with velocities less than the escape velocity of the galaxy will eventually turn around, fall in again, pass near the galactic center, and subsequently flow out again. Still lacking enough velocity to break free, this turn-around process will simply continue repeating itself yielding spatially-degenerate flows each with its own mean velocity and velocity distribution. Additionally, because the flow is effectively collisionless, dark matter caustics would form at each of the turn-around locations. Caustics are places in the galaxy where the axion density is locally enhanced. The outer caustics are topological spheres where the axions of a given outflow reach their maximum distance from the galactic center before falling back in. The inner caustics are rings where the axions with the most angular momentum in a given inflow reach their distance of closest approach to the galactic center before moving back out. Enumeration of late-infall flows is commonly done based on their direction, either falling into or out of the galaxy, and the number of times they have fallen in that direction. That is, the first infall has just entered the galaxy, the first outfall has just passed by the galactic center for the first time, the second infall has just turned around for the first time, the second outfall has just passed by the galactic center for the second time, and so on. Because relic axions would have a velocity dispersion of $\ord(10^{-17}c)$ prior to entering the galaxy,\cite{Sikivie.99} these flows are expected to have very small velocity dispersions. These small dispersions serve to keep the dark matter density within caustics finite.

It is estimated that in the neighborhood of our Sun each of the flows that has fallen into or out of the galaxy only a few times would have a density of roughly 0.01 GeV/cm$^3$ \cite{Sikivie.97}. Using 0.52 GeV/cm$^3$ as an estimate of the average local dark matter density \cite{Gates,Salucci.10}, each of these flows would contribute about 2$\%$ towards the local density. Further, observational data from the IRAS map of the galactic disk suggests that the solar system may in fact reside in the vicinity of a caustic ring. This caustic would be generated by the flows that are falling into and out of the Milky Way for the fifth time. These two flows are predicted to respectively contribute 0.95 GeV/cm$^3$ and 0.08 GeV/cm$^3$ to the local dark matter density. Possessing a density exceeding that of the local average puts a premium value on the higher density flow, referred to as the ``Big Flow.'' The expected velocity relative to the sun and velocity dispersion of the Big Flow are $v \simeq 480$ km/s and $\delta v \lesssim 53$ m/s respectively \cite{Sikivie.03}.

Flows of non-virialized dark matter may also be present due to tidal effects within the Milky Way halo. As a part of large structure formation, \lcdm\ allows for subhalos to form when smaller structures are incorporated during accretion. Tidal forces of the galaxy acting on a subhalo could cause dark matter to bleed off, forming a tidal stream of dark matter. Freese \emph{et al}. \cite{Freese.04,Freese.05} suggest that such a stram originating from the Sagittarius dwarf galaxy would posses a velocity relative to the Sun of $v \simeq 300$ km/s and a velocity dispersion of $\delta v \simeq 20$ km/s at the location of our solar system.

Considering again 3.5 $\mu$eV axions, the spectral widths of signals from the Big Flow and the Sagittarius tidal flow would be $\Delta f \lesssim$ 479 mHz and $\Delta f \simeq$ 113 Hz respectively. Sensitivity to these signals can be increased by selecting a resolution for the measured power spectrum that more closely matches the expected signal widths. Reducing the bin width of the power spectrum has the effect of reducing the noise power seen in each bin. This noise reduction yields an increase in the signal to noise ratio (SNR), provided that the spectral bin width remains at least as large as the width of the axion signal. The High Resolution (HR) channel does just this, making it more apt to identify these signals. With a native bin width of 42 mHz and the capability of generating coarser resolution spectra during processing, the HR channel is sensitive to a range of models for non-virialized axions.

\section{\label{sec:ADMX}ADMX Overview}

\begin{figure}
 \begin{center}
  \epsfig{file=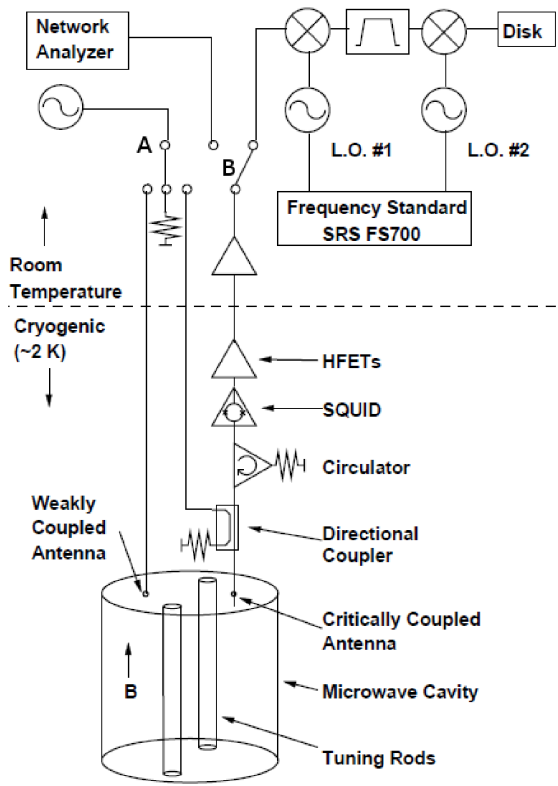,width=3in}
  \caption{\label{fig:chain}A schematic of the experimental apparatus for ADMX, showing the cavity, receiver chain, and ancillary electronics. During each measurement, switches A and B are arranged as shown. However between measurements and when performing diagnostics, power may be sent to either the major or minor port of the cavity via adjustment of Switch A, while power may be read out by a network analyzer prior to mixing via adjustment of switch B.}
 \end{center}
\end{figure}

The experimental apparatus for ADMX is shown in Fig.~\ref{fig:chain} \cite{Receiver.11,Peng.00}. At its core is a right circular cylindrical resonant cavity kept at a temperature of $T\approx2$ K. Two metal tuning rods run along the full length of the interior of the cavity and are situated parallel to the cavity axis. Horizontal translation of these rods permits the resonant frequency of the cavity to be tuned over a range of values. Both the cavity walls and tuning rods consist of high-purity copper electroplated on a stainless steel structure. Accounting for the volume of the rods, the remaining usable volume of the cavity is 140 L.

The cavity sits in a uniform 7.9 T magnetic field generated by a superconducting solenoid which is coaxial with the cavity. The magnetic field acts as a virtual photon for the interaction governed by Eq.~\ref{eq:lagg} \cite{Sikivie.83,Sikivie.85}. As such, when an axion of mass $m_a$ decays in the cavity, it would generate a single photon possessing the entirety of the axion's energy. The frequency of this photon is given by
\begin{equation}
 \label{eq:E}
 f = \frac{E_a}{h} = \frac{1}{h}\left(m_ac^2+\frac{1}{2}m_av^2\right),
\end{equation}
where $E_a$ is the energy of the axion, $h$ is Plank's constant, and $v$ is the axion's velocity relative to the detector. The rate of axion to photon conversion is enhanced by the quality factor, $Q$, of the cavity when the frequency of the decay photon is on resonance with the cavity mode.

Resonant conversion of axions to photons would result in power, $P_a$, being deposited in the cavity, half of which is coupled out and brought to the front end of the amplifier chain. The other half of the deposited power is dissipated in the cavity walls. In terms of the physical parameters involved in the conversion, the power deposited is\cite{Sikivie.83,Sikivie.85}
\begin{equation}
 \label{eq:Pa1}
 P_a = g_{a\y\y}^2\frac{VB^2\rho_aC}{m_a}\min(Q,Q_a),
\end{equation}
where $V$ is the cavity volume, $B$ is the strength of the applied magnetic, $\rho_a$ is the local axion density, and \cite{Krauss.85}
\begin{equation}
 \label{eq:Q_a}
  Q_a \equiv \frac{m_a c^2}{\Delta E} \sim 10^6
\end{equation}
is the quality factor of the energy distribution for virialized axions. Further, the strength of the axion to two photon coupling in Eq.~\ref{eq:Pa1} is $g_{a\y\y}$, which is given by
\begin{equation}
 \label{eq:gagg}
 g_{a\y\y} = \frac{\alpha g_\y}{\pi f_a}.
\end{equation}
Lastly, $C$ is a mode dependent form factor of order 1. Physically, $C$ is a unitless measure of the overlap between the electric field of the resonant mode and the applied magnetic field. This overlap is quantified as
\begin{equation}
 \label{eq:C}
 C = \frac{\left|\int_V\textbf{E}\cdot\textbf{B}_0 d^3x\right|^2}{VB_0^2\int_V\left|\textbf{E}\right|^2 d^3x}
\end{equation}
where $\mathbf{E}$ is the electric field of the resonant mode and $\mathbf{B}$ is the applied magnetic field. Equation~\ref{eq:C} has been solved numerically for various rod configurations spanning the full range motion of the rods. The form factors for rod positions not explicitly included in the calculated form factor map are interpolated from neighboring values. ADMX maximizes $C$ by using the TM$_{010}$ mode yielding calculated values for $C$ ranging from 0.41 to 0.61.

The cavity is critically coupled to the pick-up antenna resulting in half of the resonant power being sent to the RF electronics while the other half is dissipated in the cavity walls. The first stage amplifier is a DC superconducting quantum interference device (SQUID) coupled to a Nb microstrip coil, which permits high-gain, low-noise amplification over a range of operational frequencies.\cite{Receiver.11}

The microstrip SQUID amplifier (MSA) used by ADMX provids a gain of $\sim$10--12 dB with a noise temperature of roughly half the physical temperature ($T_{N,MSA} \approx 1$ K). Following the MSA are 2 GaAs HFET amplifiers constructed by the National Radio Astronomy Observatory\cite{Bradley.03,Daw.97}, each having a noise temperature of $T_{N,HFET} \approx 4$ K and a gain of 17 dB. The contributions to the system noise temperature from each HFET are suppressed by the preceding amplification. This results in contributions of 0.4 K for the first HFET and 0.008 K for the second. Thus the system noise temperature, being the sum of the physical temperature and the effective noise temperatures of the MSA and both HFETs, is taken to be $T_{N,Sys} \approx 3.4$ K.

A magnet quench in late 2009 critically damaged the SQUID in the MSA, rendering it useless. Because a replacement MSA was not immediately  available, data taken from February 2010 through April 2010 bypassed the MSA making the first HFET the first-stage amplifier instead. Lacking the noise suppression provided by the MSA, the system noise temperature for this set of data is $T_N \approx 6$ K. Even without the MSA, all amplifiers following the HFETs provide negligible contributions to the system noise. Additionally, when performing the MSA bypass, it was noticed that there was a poor electrical connection between the walls and the lid of the cavity. This issue had caused the data up to this point to suffer from low $Q$ values, typically 10,000--20,000. The connection was repaired, increasing the measured $Q$ to the 50,000 to 80,000 range. Overall, increasing the $Q$ more than compensates for the increased noise temperature, leading to noticeably better density limits for the 2010 data.

The room temperature portion of the receiver chain consists of 4 commercial microwave amplifiers at 30 dB of gain each, 2 image-reject mixers which mix the signal down to a center frequency of 10.7 MHz then again down to 35 kHz, and a crystal bandpass filter with a spectral bandwidth of 30 kHz. The losses in the receiver chain are 7 dB for each of the two mixers, 3 dB for the crystal filter, and 6 dB for the cables in the transmission line. The total amplification of the signal prior to digitization is thus 141 dB prior to February 2010 and 131 dB afterward\cite{Receiver.11}.

After amplification, the MR and HR channels diverge for digitization. The MR channel measures the incoming signal in 80 second blocks wherein 10,000 individual 8-ms-long measurements are recorded. These measurements, taken at a rate of 100,000 samples/second, are converted to power spectra via fast Fourier Transform (FFT) and averaged into a single power spectrum with a bin width of 125 Hz and a bandwidth of 50 kHz. In this same time, the HR channel makes three separate measurements of the incoming signal. Each is at a rate of 80,000 samples/second for 23.8 seconds. The three time traces are each saved directly to disk at this time without any additional processing or averaging. For both channels, all relevant state data for the experiment, such as the time stamp, physical temperature, magnetic field strength, Q, etc, are recorded immediately prior to each measurement and are saved in the header of each data file.

Finally, the tuning rods are moved such that the resonant frequency of the cavity is stepped by about 2 kHz, and state data for the next measurement are recorded. Movement of the tuning rods and the acquisition of the state data collectively take about 20 seconds resulting in a time between measurements of about 100 seconds. The data considered here were taken between March 31, 2009 and April 1, 2010 and cover a frequency range of 812--852 MHz and 858--892 MHz. Note that ADMX still collected data for the HR channel prior to March 31, 2009. Unfortunately, the raw signal voltage versus time data for these measurements were not saved. Instead, the three measurements made for each rod configuration were converted to power spectra and averaged before being saved to disk. This precludes their use in the present analysis which leads to a frequency gap from 852 to 858 MHz. An analysis of these data with a 10.8 Hz frequency resolution can be found in Ref. \cite{Hoskins.11}.

\section{\label{sec:res}Resolution Selection}

As mentioned before, there is a direct relationship between the velocity dispersion of a population of axions and the width of the signal they generate in the ADMX detector. Matching these properties is crucial to maximizing ADMX's detection capabilities. While the native bin width of the HR channel is 42 mHz, coarser resolutions can and should be generated from the time traces in order to fill the resolution gap between the HR and the MR channels. An analysis using a bin width of 10.8 Hz has already been performed\cite{Hoskins.11}, although this still leaves a gap for bin widths of $\lesssim$ 1 Hz. To rectify this, power spectra with integer multiples of the base resolution were produced resulting in bin widths of 84 mHz, 168 mHz, 546 mHz, and 1.09 Hz for $n =$ 2, 4, 13, and 26 respectively.

To obtain the maximum dispersion for which a particular resolution is sensitive, the spectral width is set equal to the bin width ($b = \Delta f$) and Eq.~\ref{eq:ffrac} rewritten for $\delta v$, yielding
\begin{equation}
 \label{eq:disp}
 \delta v = \frac{c^2}{fv} \left(\frac{b}{2}\right).
\end{equation}
For a signal in the middle of our search range, $f=850$ MHz, and a mean flow velocity of $v\approx 300$ km/s, Eq.~\ref{eq:disp} then yields maximum dispersions in order of ascending $n$ of 15, 30, 96, and 192 m/s. While these resolutions place the Sagittarius tidal flow outside the scope of the present analysis, the Big Flow remains in consideration and would be most prominent in the $n = 13$ spectra.

Selecting coarser resolutions can be done either in the frequency domain, by adding neighboring bins in the power spectrum, or in the time domain, by subdividing the time trace into equal length sections data prior to applying the FFT. Note that resolution selection in the time domain requires that the power spectra generated for each subsection be co-added to maintain the overall SNR. Previous analyses\cite{Duffy.06,Hoskins.11} used the former method while the present analysis uses the latter. The SNRs for both methods are identical for signals that remain coherent over the entire integration time of the measurement. The same can not be said of a signal that becomes decoherent, say by means of a drifting phase, part way through integration. A signal of this sort will suffer from losses due to its partial decoherence when a single FFT is applied to the entire data stream. In contrast, subdividing the time trace allows coherence to be maintained within each section while not requiring that it be maintained from one section to the next.

This approach was verified via numerical simulation, whereby a decoherent signal was generated and processed using each of the above methods. The length of the generated data stream was selected to match that of the HR channel (23.8-second integration time) and the desired bin width was set to 420 mHz (i.e., $n = 10$). The signal was modeled as a sinusoid with a variable phase and a coherence time of roughly 2.38 seconds. The actual times at which the phase changes occurred were determined using a Gaussian distribution about the coherence time. At these times, the phase was abruptly changed to a new randomly determined value. To be consistent with the noise characteristics of the HR channel, the signal was added to randomly generated background of exponential noise (SNR=50). Each method of creating coarser resolution spectra was tested 10,000 times and the average results were compared with the output being normalized to that of a perfectly coherent signal. Resolution selection in the frequency domain exhibited an average loss of 38$\%$, while selection in the time domain lost only 5$\%$. For reference, the power of a perfectly decoherent signal would be reduced by a factor of $1/\sqrt{n}$, or a loss of 68.4$\%$. Clearly, there is a benefit to ensuring that the integration time remains shorter than the coherence time.

Due to a hardware limitation, the coherence time of the ADMX data stream is shorter than the integration time of the HR channel. The frequencies of the mixers used in the receiver chain are set by an external frequency standard (SRS FS700), which has a short term stability of 10$^{-10}$\cite{FS700}. This instability is negligible for the MR channel, but not for the HR channel. The phase noise in the first mixer caused by jitter in the frequency standard limits the narrowest bin width used for the HR channel to 84 mHz, or that of the $n = 2$ spectra. While the $n = 1$ spectra are still viable for the purposes of verifying noise statistics and identifying data corruption from external sources, the actual exclusion limits presented later will only consider the $n =$ 2, 4, 13, and 26 spectra.

Finally, assuming that the measured signal actually remains coherent over the entire integration time, it is prudent to confirm that the SNR is indeed the same for both methods. Consider an axion signal with power $P_a$, as given by Eq.~\ref{eq:Pa1}, which is spread equally over $n$ bins in the 42-mHz-resolution spectrum. In the frequency domain, the total power is recovered by adding these $n$ bins together, increasing the measured power by a factor of $n$. Because the noise power is added in quadrature, thus only increasing by a factor of $\sqrt{n}$, the net gain in the SNR is that of a factor of $\sqrt{n}$. In the time domain, each subspectrum would contain the entire signal power in a single bin, yielding a factor of $n$ increase in measured power, and when the subspectra are recombined that power would add constructively, yielding another factor of $n$ increase. The rms noise power of each subspectrum is proportional to its bin width, which increases the noise by a factor of $n$, and the noise is again added in quadrature during recombination, for another factor of $\sqrt{n}$ increase. As was seen in the frequency domain, the net effect is that the SNR is increased by a factor of $\sqrt{n}$.

\section{\label{sec:mod}Signal Modulation}

The populations of axions considered by ADMX are assumed to have average velocities, velocity dispersions, and densities which vary on timescales that are much larger than the cumulative integration time of the experiment. As such, flows are considered to be steady and for each flow these properties are all taken to be constant. Despite this, these effectively constant properties will still vary relative to the detector. The rotational and orbital motions of the Earth will cause noticeable, but predictable, variations in the signals produced in the ADMX detector\cite{Ling.04,Turner.90}. Specifically, both would cause a modulation of the axion signal over timescales shorter than the lifetime of the experiment, and must therefore be scrutinized in some detail.

Similar to spectral broadening, signal modulation can be examined by using Eq.~\ref{eq:ffrac}. In this case $\delta v$ is taken to be the difference between the present and time-averaged values of the mean velocity of a flow relative to the detector. To quantify the typical scales of signal modulation caused by the Earth's motion, consider a signal in the middle of our data set, $f=850$ MHz, and a mean flow velocity of $v\approx 300$ km/s. At the time of data acquisition, the detector was located a few hundred feet above sea level at a latitude of 37.68$^\circ$ North. Moving 2$\pi$ radians in one sidereal day at this location yields an amplitude for the rotational component of $\delta v$ of 368 m/s. The amplitude of the diurnal signal modulation is then obtained by substituting these values into Eq.~\ref{eq:ffrac} resulting in
\begin{equation}
 \label{eq:dfr}
 \delta f_r = 1.04 \tn{ Hz}.
% \delta f_r = \frac{(850\tn{MHz})(300 \tn{km/s})(368 \tn{m/s})}{c^2} = 1.04 \tn{Hz}.
\end{equation}
A similar calculation is done for the amplitude of the annual modulation, using 29.8 m/s as the amplitude of the orbital component of $\delta v$, which gives
\begin{equation}
 \label{eq:dfo}
 \delta f_o = 84.5 \tn{ Hz}.
% \delta f_o = \frac{(850\tn{MHz})(300 \tn{km/s})(29.8 \tn{km/s})}{c^2} = 84.5 \tn{Hz}.
\end{equation}
Because the phases of the rotational and orbital velocity oscillations are both unknown, the signal could actually drift by as much as twice these values, in either direction, over the course of one half period (i.e., $\pm 2 \delta f$).

It is convenient to look at the short-term effects of both modulations, thereby allowing them to be compared using equal timescales. Because the velocity oscillations are effectively sinusoidal, the maximum rate of change for times much smaller than the period, $T$, is given by
\begin{equation}
 \label{eq:vdot}
 \dot{\delta v}_{max}= \frac{2\pi A}{T},
\end{equation}
where A is the amplitude of the velocity oscillation. For $t=12$~hours, or one half rotation, $t$ is much less than the orbital period. This permits the use of $\delta v \approx \dot{\delta v}t$ in Eq.~\ref{eq:ffrac}, yielding
\begin{equation}
 \label{eq:maxmodO}
 \delta f_{o,\ 12\ \tn{hours}} =  0.73 \tn{ Hz}.
% \delta f_{12\ \tn{hours}} = 2\pi(84.5\tn{Hz})\frac{12\tn{hours}}{1 \tn{year}} = 0.73 \tn{Hz}.
\end{equation}

Both of these amplitudes represent the worst-case scenarios regarding flow orientation, parallel to the equator for diurnal modulation and parallel to the ecliptic for annual modulation. Note that both conditions can not simultaneously be true at all times, but these values are still reasonable measures of their maximum possible amplitudes. Specific knowledge of the directions of axion flows relative to the motion of the Earth would be needed to refine these estimates. Additional sources of modulation include motion about the Earth-Moon barycenter as well as motion about the galactic center. Both are considered negligible as the former has a trivial amplitude and the latter is of a timescale significantly longer than the lifetime of the experiment.

On the timescale of one 23.8-second measurement, diurnal and annual modulations only shift the frequency of the signal by 1.8 mHz and 0.4 mHz respectively. Being much smaller than the native 42 mHz bin width of the HR channel, any spectral broadening caused by signal modulation during a single measurement is considered to be inconsequential. However, signal modulation, even over times of only 10 to 20 minutes, is sufficient to move an actual axion signal from one frequency bin to another. Thus, unlike the MR channel, which calculates the power measured at each frequency by taking a weighted average of the power measured from all spectra containing that frequency, the HR channel must consider each spectrum individually. The implications of this fact are discussed further in section \ref{sec:lim} of the paper.

\section{\label{sec:noise}Noise Properties}

\begin{figure}
 \begin{center}
  \epsfig{file=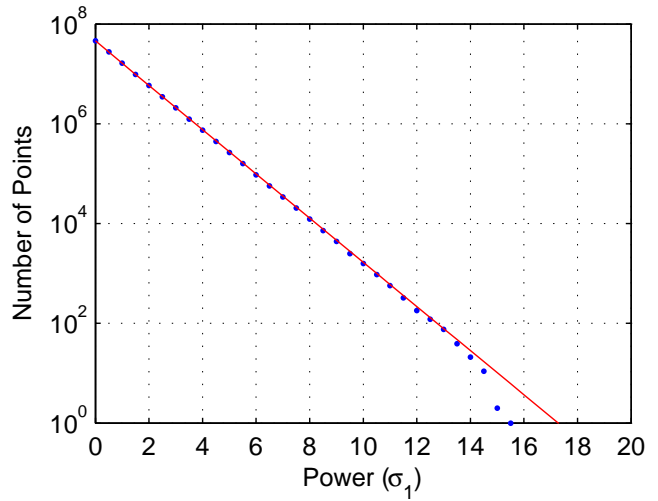,width=3.4in}
  \caption{\label{fig:noise1}Power distribution for $n = 1$ using a subset of the HR data consisting of 200 spectra.}
 \end{center}
\end{figure}

It has been established that when no averaging is performed the noise distribution for the HR channel follows exponential statistics \cite{Duffy.06}, with the standard deviation, $\sigma$, of the distribution being set by the rms noise power of the spectrum. That is,
\begin{equation}
 \label{eq:rms}
 \sigma = P_N = T_N k_b b,
\end{equation}
where $P_N$ is the rms noise power, $T_N$ is the system noise temperature, $k_b$ is Boltzmann's constant, and $b$ is the spectral bin width. Thus, the probability, $P$, of measuring a noise power of $p$ at a given frequency is determined by
\begin{equation}
 \label{eq:pdf1}
 dP = \frac{1}{\sigma}~e^{(-p/\sigma)}dp.
\end{equation}
Not only should the $n = 1$ spectra follow these statistics, so too should all subspectra generated for $n > 1$. The only difference is that the rms noise power is a function of $b$, thus requiring that $\sigma$ instead be denoted as $\sigma_n$. A histogram of the power distribution for $n = 1$ data is shown on a semilog plot in Fig.~\ref{fig:noise1} along with a line showing the expected distribution. Examination of the power distributions for the $n > 1$ subspectra revealed that they too follow the expected exponential behavior. When the spectra are plotted this way, the slope of the best fit line is actually $-1/\sigma_n$, permitting each subspectrum to be renormalized to units of $\sigma_n$.

\begin{figure}
 \begin{center}
  \epsfig{file=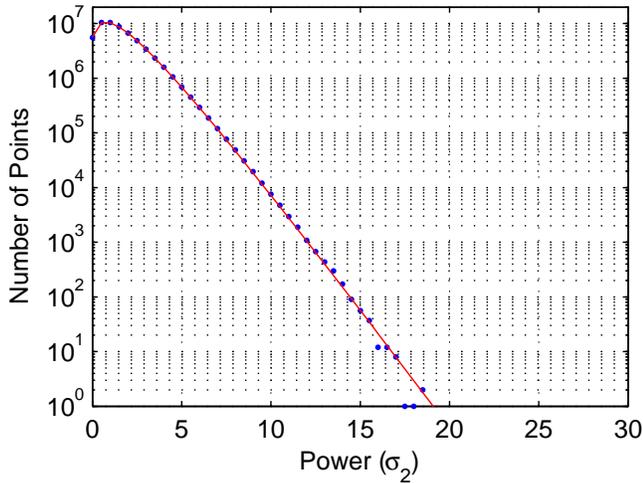,width=3.4in}
  \caption{\label{fig:noise2}Power distribution for $n = 2$ using the same subset of data shown in Fig.~\ref{fig:noise1}.}
 \end{center}
\end{figure}

\begin{figure}
 \begin{center}
  \epsfig{file=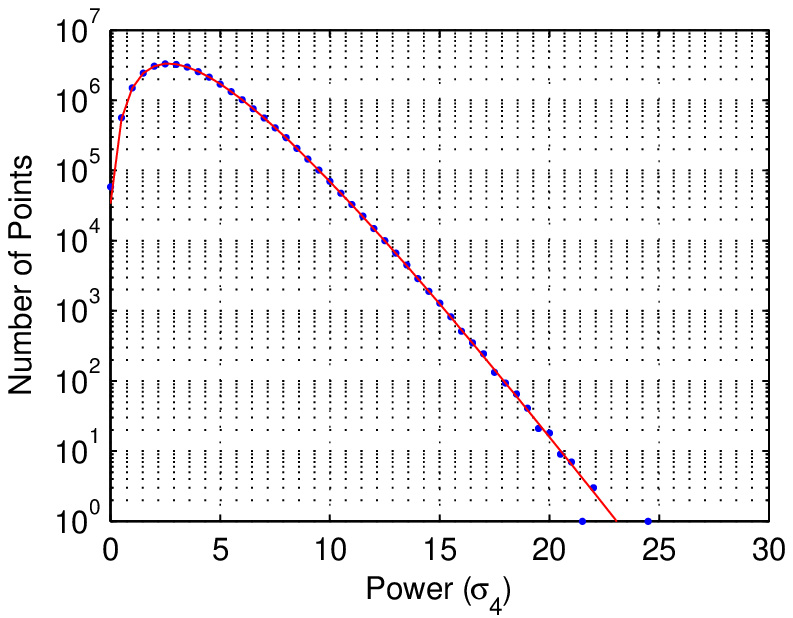,width=3.4in}
  \caption{\label{fig:noise4}Power distribution for $n = 4$ using the same subset of data shown in Fig.~\ref{fig:noise1}.}
 \end{center}
\end{figure}

\begin{figure}
 \begin{center}
  \epsfig{file=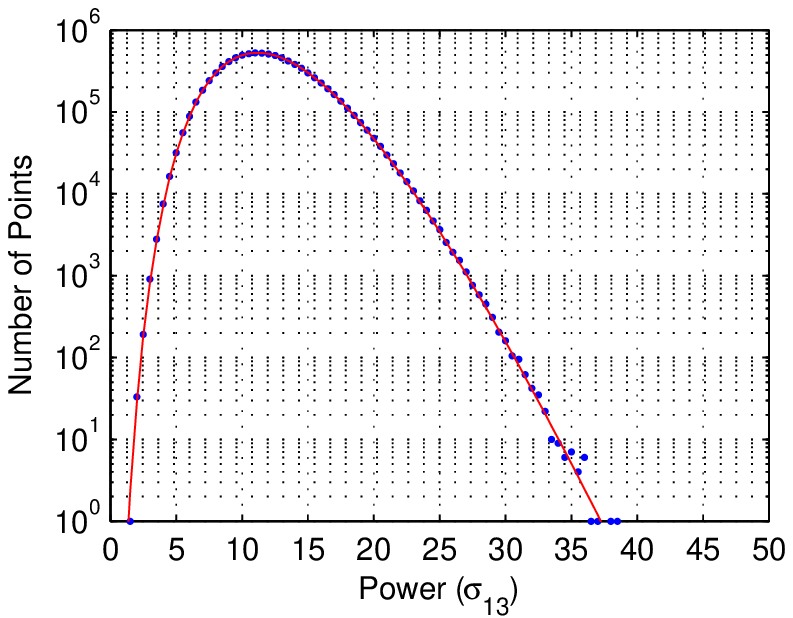,width=3.4in}
  \caption{\label{fig:noise13}Power distribution for $n = 13$ using the same subset of data shown in Fig.~\ref{fig:noise1}.}
 \end{center}
\end{figure}

\begin{figure}
 \begin{center}
  \epsfig{file=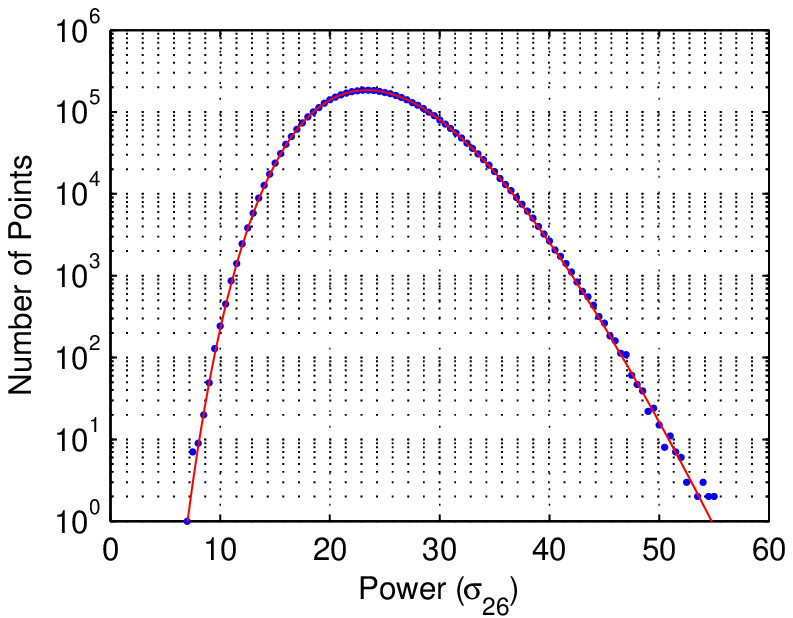,width=3.4in}
  \caption{\label{fig:noise26}Power distribution for $n = 26$ using the same subset of data shown in Fig.~\ref{fig:noise1}.}
 \end{center}
\end{figure}

Because the subspectra for a given time trace must be recombined for each resolution before searching for an axion signal, the final noise distribution deviates noticeably from Eq.~\ref{eq:pdf1}. Instead, a convolution of $n$ exponential distributions, each with rms noise power $\sigma_n$, yields the actual distribution for the final spectra at each resolution. The general form of this expression is given by
\begin{equation}
 \label{eq:pdfn}
 dP_n = \frac{p^{(n-1)}}{(n-1)! \sigma^n_n}~e^{(-p/\sigma_n)} dp.
\end{equation}
The noise distributions for a subset of the recombined $n =$ 2, 4, 13, and 26 data sets are shown in Figs.~\ref{fig:noise2}, \ref{fig:noise4}, \ref{fig:noise13}, and \ref{fig:noise26} respectively. The solid lines in these plots are the expected distributions as determined by Eq.~\ref{eq:pdfn}. Good agreement is seen between the actual and expected distributions for each resolution, as is the trend towards a Gaussian in accordance with the central limit theorem.

\section{\label{sec:errors}Systematic Errors}

\begin{figure}
 \begin{center}
  \epsfig{file=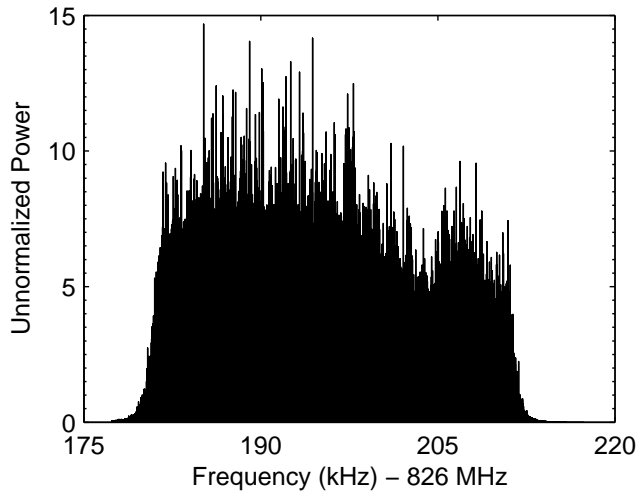,width=3.4in}
  \caption{\label{fig:raw}A typical example of a power spectrum from the HR channel. The roll-off at the edges is due to the bandwidth of the HR spectrum being larger than that of the crystal filter. The waviness in the central portion of the spectrum stems directly from the spectral shape of the crystal filter. A reference spectrum showing this shape can be seen in Fig.~\ref{fig:xtal}.}
 \end{center}
\end{figure}

Before they can be searched for axion signals, all data files must be corrected for systematic errors. Specifically, the spectral shape imparted to the data by the receiver chain must be removed. Because the crystal filter produces a distinct spectral shape, shown in Fig. ~\ref{fig:raw}, it is considered separately from the net effect of the other electronics in the receiver chain. The latter error, referred to as the receiver response, varies widely over the data set as it is highly frequency dependent. Corrections for these effects are applied to each subspectrum at each resolution, after an initial estimate of $\sigma_n$ has been made. Once corrected, the actual value of $\sigma_n$ can be determined and the subspectra can be renormalized and recombined. Note also that the correction procedures described below are still applied to the $n = 1$ data set despite its exclusion from consideration regarding limits. To confirm noise statistics and identify data contamination properly, an identical treatment of the $n = 1$ data is required.

\begin{figure}
 \begin{center}
  \epsfig{file=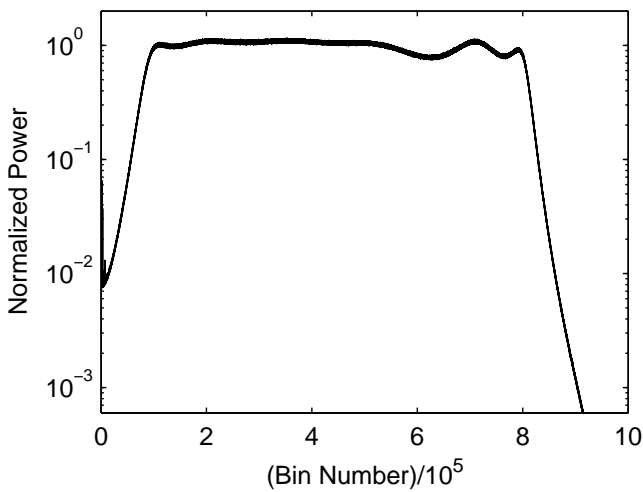,width=3.4in}
  \caption{\label{fig:xtal}A reference spectrum showing the spectral shape of the crystal filter. This reference was created by averaging about 10,000 individual spectra which cover a range of frequencies. }
 \end{center}
\end{figure}

For each resolution, approximately 10,000 spectra are normalized to a mean value of 1 and are averaged to create a reference spectrum for the shape of the crystal filter. One such reference spectrum is shown in Fig.~\ref{fig:xtal}. Using spectra centered on a wide range of frequencies ensures that the frequency dependent receiver response is sufficiently averaged so as not to bias the reference spectrum. New reference files were created periodically to account for long term variations in the spectral shape of the crystal filter. After each subspectrum is divided by the appropriate reference spectrum, its first and last 20$\%$ are cropped, leaving only the section corresponding to the more or less flat portion of the usable bandwidth of the crystal filter. Any large scale structure remaining in the now 24-kHz wide subspectrum is attributed to the receiver response.

\begin{figure}
 \begin{center}
  \epsfig{file=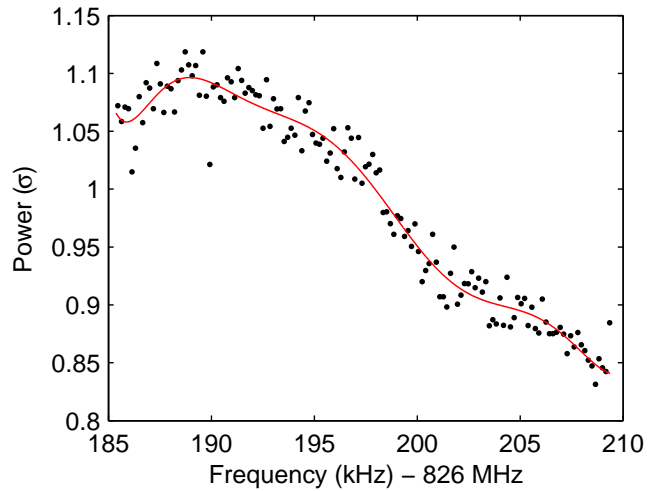,width=3.4in}
  \caption{\label{fig:poly9}A polynomial of degree 9 is fit to a cropped coarse resolution spectrum following the removal of the crystal filter shape. This fit is the receiver response function which is used to remove any residual shape from the truncated spectrum.}
 \end{center}
\end{figure}

The first step in correcting for the receiver response is to create a coarse resolution spectrum, $b = \ord$(100 Hz), for each subspectrum by averaging neighboring bins. The exponential nature of the noise in the HR channel is well averaged at this resolution which permits the spectrum to be fit with a polynomial of degree 9, as seen in Fig.~\ref{fig:poly9}. Data points with higher power than the end of the exponential distribution ($p \gtrsim 18\sigma$) are ignored when creating the coarse spectrum so as not to bias the fit. The reduced chi-squared, $\chi_r^2$, of the fit is then calculated and used as a measure of how well the fit matches the receiver response. A subspectrum with a $\chi_r^2$ outside the range 0.8 -- 1.2 is considered to have a poor fit for the receiver response. Any time trace that generates at least one subspectrum with an inadequate $\chi_r^2$ was cut from the data set. Finally, the polynomial is divided out of the cropped subspectrum leaving it spectrally flat.

\begin{figure}
 \begin{center}
  \epsfig{file=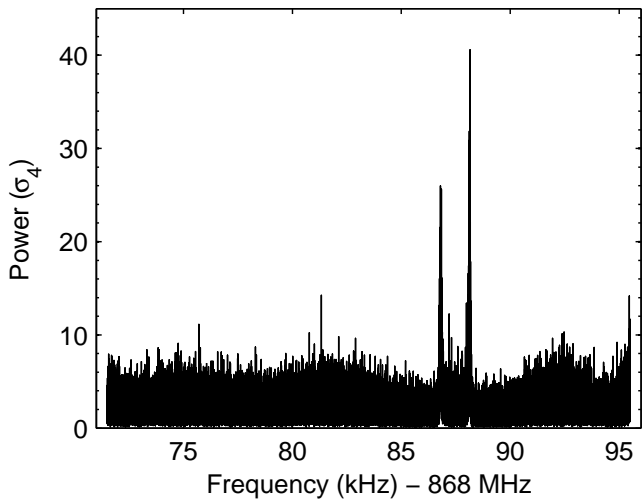,width=3.4in}
  \caption{\label{fig:radio}A strong external radio signal can be seen in the latter half of this spectrum. Such a signal would dwarf any axion signal rendering the spectrum useless.}
 \end{center}
\end{figure}

\begin{figure}
 \begin{center}
  \epsfig{file=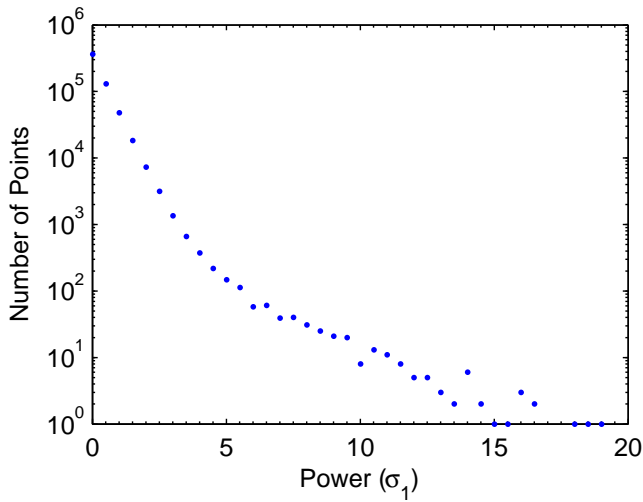,width=3.4in}
  \caption{\label{fig:RadioDist}The radio signal shown in Fig. ~\ref{fig:radio} is identified during processing by its $n = 1$ power distribution. At high powers, this distribution exhibits large deviations from the exponential behavior.}
 \end{center}
\end{figure}

Once the systematic errors have been corrected, the data are checked for contamination from external sources of non-statistical noise. Excess power in the cavity of this nature would cause the noise distribution to deviate from its expected shape. If non-statistical noise is present at one resolution it will be present for all other resolutions. Consequently, only the $n = 1$ data are used to check for its presence, as checking each resolution would be redundant. Because the $n = 1$ noise distribution would be linear on a semilog plot, deviations are relatively easy to identify. An example of this can be seen in Figs.~\ref{fig:radio} and \ref{fig:RadioDist} which respectively show the power spectrum and noise distribution for a spectrum containing an external radio signal. An improperly shielded RF cable led to this sort of signal leaking into a large number of files between February 19, 2010 and March 1, 2010. All affected files were cut. Once discovered, the cable was promptly replaced, with no further radio contamination occurring thereafter.

In total, 274,834 time traces were recorded and corrected for systematic errors between March 31, 2009 and April, 1 2010. Of these, 246,556 time traces passed all data quality cuts and were subsequently searched for axion signals.

\section{\label{sec:search}Search Procedure}

Recall that on timescales considered for ADMX, each flow is considered to be steady and ever-present. Thus, a measurement of any real axion signal carries with it the expectation of repeatability. The search procedure must still account for signal modulation, but this only requires that candidate signals in one spectrum be measured at nearby frequencies in spectra with overlapping frequency coverage. However, what is considered to be a ``nearby'' frequency is different for each pair of spectra, and is determined by the elapsed time between when the spectra were measured. Also recall that three time traces are taken per tuning rod configuration, the step size of the cavity tuning is about 2 kHz, and the bandwidth of each spectrum is 24 kHz. Further, it is not uncommon for frequencies to be examined again weeks later. This all results in the frequency coverage of each spectrum being overlapped by numerous other spectra which were measured over a fairly broad time span. With these considerations in mind, the procedure used for identifying potential axion signals is now described.

First, the recombined spectra were indexed chronologically, and a single file was created to hold all of the state data contained in the headers of each spectrum. Each spectrum was then searched for signals exceeding a resolution dependent candidate threshold, which is discussed below. The frequencies of all candidate signals were recorded along with the signal power and the spectrum index. Because the state data are the same for all candidates in a given spectrum, the index file permits one copy of the state data to be loaded into memory per spectrum rather than one copy per candidate in the spectrum.

Each candidate signal is then compared to all candidate signals in each subsequent spectrum for which the difference in the signal frequencies is less than 1 kHz. This cutoff for the frequency spacing serves to reduce the number of unnecessary calculations performed. The worst case estimates of both annual and diurnal modulations predict a maximum possible frequency drift which is still well below 1 kHz. The maximum expected modulation is then calculated from the difference in the time stamps of the candidates and the worst-case, short-term modulations given by Eqs.~\ref{eq:dfr} and \ref{eq:maxmodO}. Should the candidate frequencies be separated by less than this maximum modulation, they may represent the same signal and are considered to be coincident. For each candidate, the number of spectra for which there is at least one coincident data point is compared to the total number of spectra covering the appropriate nearby frequencies to determine if there is statistically significant coincidence which would warrant further consideration of the candidate.

The benchmark for what is considered statistically significant must account for the suppression of off-resonance signals which are still within the cavity bandwidth. The spectral shape of the resonant mode of the cavity is a Lorentzian given by
\begin{equation}
 \label{eq:L4}
 L(f) = \frac{1}{1+4Q^2\left(1-\frac{f}{f_0}\right)^2},
\end{equation}
where $f_0$ is the resonant frequency of the mode and $Q$ is the quality factor of the cavity. Any signal which is sufficiently far away from the center of the Lorentzian will be suppressed by a factor of $L$, which may lower the signal power below the candidacy threshold. The useful frequency range is effectively limited by the full width at half maximum ($\mbox{\em FWHM}$) of the Lorentzian, which is given by
\begin{equation}
 \label{eq:Thresh1}
 \mbox{\em FWHM} = \frac{f_0}{Q}.
\end{equation}
Any expectation that a signal will consistently be seen despite being suppressed for residing outside the $\mbox{\em FWHM}$ is unrealistic and would bias the search against seeing significant coincidence.

For a spectrum with $f_0 \approx 850$ MHz and $Q \approx 35,000$, the $\mbox{\em FWHM}$ would roughly match the 24 kHz bandwidth of the spectrum. Further, doubling the $Q$ would halve the usable frequency range. About half of the spectra have a $Q$ which exceeds 35,000, while only about 17$\%$ of the spectra have a $Q$ exceeding 70,000. Given these values, any candidate signal showing coincidence with at least 50$\%$ of the overlapping spectra is considered to have significant coincidence. Such a candidate could not be ruled out without further testing.

To select values for the candidate thresholds, two competing factors must be considered and carefully balanced. Because the limits on the local dark matter density are directly proportional to the candidate thresholds, it would seem optimal to select thresholds which are as low as possible. However, the computational time of the search grows as the square of the number of candidates, which itself grows exponentially as the thresholds are lowered. To resolve this, initial estimates of the candidate thresholds were determined by numerically integrating Eq.~\ref{eq:pdfn} to find the power at which there is roughly a 10$\%$ chance of seeing a pure noise peak in a single spectrum. These estimates were 17.6$\sigma_2$, 21.8$\sigma_4$, 36.6$\sigma_{13}$, and 55.2$\sigma_{26}$. Setting the candidate thresholds slightly lower than these estimates, at 16$\sigma_2$, 20$\sigma_4$, 34$\sigma_{13}$, and 51$\sigma_{26}$ respectively, yielded about 60,000 candidates per resolution, none of which showed significant coincidence. Upper limits on the local density of non-virialized axions were then calculated for the frequency range of 812 to 852 MHz and 858 to 892 MHz. Using these lower thresholds slightly improved the density limits while only incurring a modest increase in the computational time due to having a larger set of candidates.

\section{\label{sec:lim}Density Limits}

Because the HR spectra are not averaged together, a density limit is determined for each spectrum and a weighted average of these limits is calculated. However, before any limits are generated an adjustment is made to the candidate thresholds, which better accounts for the background noise. Additionally, the expected axion power is adjusted for losses from the cavity-antenna coupling, the Lorentzian shape of the resonant mode, and the spreading of power to other frequencies via the FFT. The first loss is simply a factor of 2, while the other losses require a bit more consideration.

Lacking a real axion signal, the candidate thresholds are treated as the sum of the noise power and the expected axion power. The noise power in this case is no longer $\sigma_n$, but is instead some value which is consistent with Eq.~\ref{eq:pdfn}. With a desired confidence level of 90$\%$, the power below which there is only a 10$\%$ chance of finding a pure noise peak must be calculated. This was done by numerically integrating Eq.~\ref{eq:pdfn} for each noise distribution. These values were subtracted from the candidate thresholds resulting in effective thresholds of 15.5$\sigma_2$, 18.3$\sigma_4$, 25.4$\sigma_{13}$, and 31.3$\sigma_{26}$.

The correction for the mode shape is complicated slightly by the Lorentzian being frequency dependent. A single value of $L$ must be chosen for each spectrum, but neither its maximum nor minimum values is truly representative of its effect. The value used for the expected axion power is chosen based on the aforementioned limitation imposed by the $\mbox{\em FWHM}$ and the fact that there is an equal probability of finding an axion signal at any frequency in the spectrum. As such, the mean value of the Lorentzian, $\overline{L}$, over the effective frequency range is used. This value is calculated analytically from Eq.~\ref{eq:L4} and is a constant $\pi/4$ for spectra where $\mbox{\em FWHM} < 24$ kHz. For spectra with wider Lorentzians it is instead given by
\begin{equation}
 \label{eq:Lbar2}
 \overline{L} = \frac{f_0}{(24\ \tn{kHz})Q}\arctan\left(\frac{(24\ \tn{kHz})Q}{f_0}\right) \geq \frac{\pi}{4}.
\end{equation}

The loss imparted by the FFT is an inherent aspect of the FFT itself. An FFT only returns an output for a select set of frequencies. Specifically, only frequencies which are integer multiples of the spectral resolution. Any power at a frequency which is not an element of this set is actually spread out over all frequencies in the set, with most of the power being split between the two closest elements. Logically, a signal offset from one of the output frequencies by $b/2$ will contribute equally to its closest neighbors; however, each will only receive 40.5$\%$ of the total signal power. The rest of the signal power is distributed over the remainder of the FFT output. At a given integer multiple of $b$, $f_j=jb$, the power contribution from a signal at $f=b(j+\Delta)$ is given by
\begin{equation}
 \label{eq:Pj}
  P_j=\left(\frac{\sin(\pi\Delta)}{\pi\Delta}\right)^2.
\end{equation}
For any offset such that $\left|\Delta\right| > 0.5$, the signal would no longer be seen at $f_j$, as the majority of the power would then be closer to a different element of the FFT output. Because an axion signal has an equal probability of being at any frequency, the fraction of the power contributed at $f_j$ is taken as the mean value of Eq.~\ref{eq:Pj} over a single frequency bin, that is for $\Delta$ ranging from -0.5 to 0.5. This fraction is resolution independent and is given by
\begin{equation}
 \label{eq:Pj1}
  M = \int_{-1/2}^{1/2}\left(\frac{\sin\left(\pi \Delta\right)}{\pi \Delta}\right)^2 d\Delta = 0.774.
\end{equation}

The expected axion power from Eq. ~\ref{eq:Pa1} was adjusted for these losses and equated to the effective thresholds. Including the factor of $n$ associated with the recombination of subspectra and solving for $\rho$, this relation is given by
\begin{equation}
 \label{eq:rhoLim}
 \rho = T_n \sigma_n \times\left(\frac{m_a}{g_{a\y\y}^2VB_0^2CQ}\right)\left(\frac{2}{0.774\overline{L} n}\right),
\end{equation}
where the $T_n$ are the effective thresholds in units of $\sigma_n$. Given that after recombination the rms noise power for each spectrum is $\sqrt{n}\sigma_n$, the uncertainty in $\rho$ would then be
\begin{equation}
 \label{eq:rhoErr}
 \delta = \sigma_n \times\left(\frac{m_a}{g_{a\y\y}^2VB_0^2CQ}\right)\left(\frac{2}{0.774\overline{L} \sqrt{n}}\right).
\end{equation}
To place limits on $\rho$, an assumption had to be made for the value of \gagg. As such, the values for both the KSVZ and DFSZ models were used, thus producing two sets of limits. From Eqs. \ref{eq:rhoLim} and \ref{eq:rhoErr}, it is clear that both the limits and the uncertainties for the KSVZ and DFSZ models differ only in magnitude. Scaling by
\begin{equation}
 \label{eq:gagRatio}
 \left(\frac{g_{\rm{KSVZ}}}{g_{\rm{DFSZ}}}\right)^2 = 7.26
\end{equation}
is all that is needed to convert from KSVZ results to DFSZ results. Similarly, limits for $n > 2$ can be obtained from the limits for $n = 2$ via scaling by $T_n/T_2$. Equations \ref{eq:rhoLim} and \ref{eq:rhoErr} were then evaluated for each spectrum, using $g_{\rm{KSVZ}}$ and $n = 2$.

Finally, a weighted average of the limits was taken for each 1 MHz wide section of the search range, with each spectrum contributing to all sections for which it has a frequency overlap. This average was computed according to
\begin{equation}
 \label{eq:FinalLim}
 \rho_{final} = \sum_i \frac{\delta_m^2 \rho_i}{\delta_i^2},
\end{equation}
where $\delta_m$ is defined by
\begin{equation}
 \label{eq:delta_m}
 \frac{1}{\delta_m^2} \equiv \sum_j \frac{1}{\delta_j^2}.
\end{equation}
Due to this weighting, those spectra with a low $Q$ or a high $\sigma_n$ contribute less to the density limits than do spectra with respectively higher or lower values.

\begin{figure}
 \begin{center}
  \epsfig{file=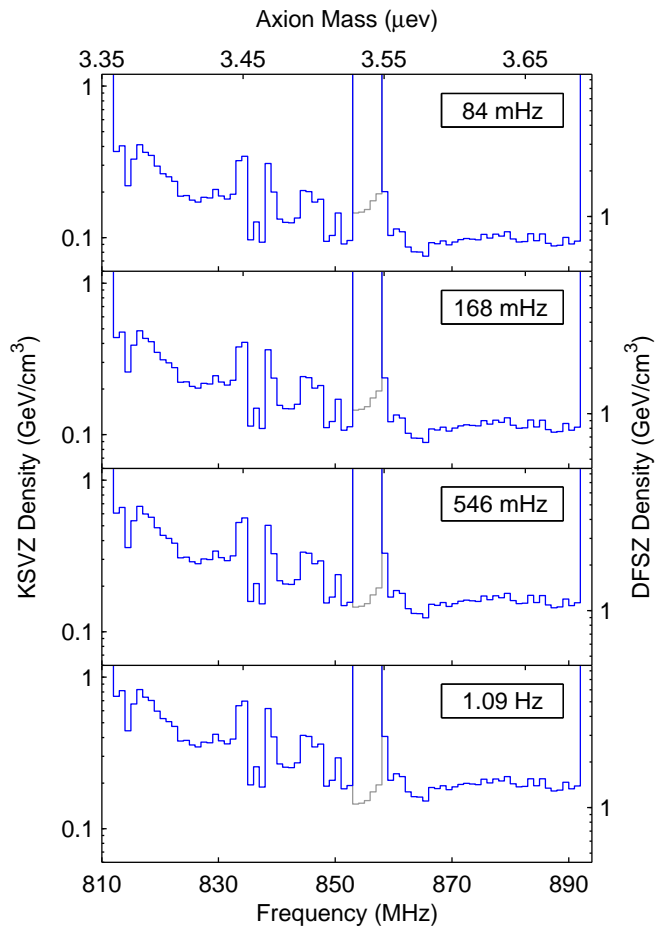,width=3.4in}
  \caption{\label{fig:limits}Exclusion limits (90$\%$ CL) placed on the local density of non-virialized axions at each resolution. Densities for KSVZ axions are shown on the left axis while those for DFSZ axions are shown on the right. The density limits shown from 853 through 857 MHz are the limits published in \cite{Hoskins.11} (shown in grey), which have a resolution of 10.8 Hz in each plot.}
 \end{center}
\end{figure}

These limits were then scaled to obtain a complete set of final limits covering both axion models at each resolution. Table~\ref{tbl:density} contains a summary of the important parameters associated with each resolution. Lastly, the final limits for this analysis are shown in Fig.~\ref{fig:limits} as are a subset of the limits obtained in \cite{Hoskins.11}. The limits from the earlier analysis are shown as a grey line between 852 and 858 MHz, covering the frequency gap for the present analysis.
\begin{table}
 \begin{center}
  \caption{Parameters associated with each resolution. Critical to the interpretation of the limits are the spectral resolution and the maximum velocity dispersions. Also included are $n$, the effective thresholds, and the scale factors.}\label{tbl:density}
  \begin{tabular*}{3.4in}{S[table-format=4.0] @{\extracolsep{\fill}} S[table-format=3.0] S[table-format=2.0] S[table-format=2.1] S[table-format=1.2]}
%  \begin{tabular}{S[table-format=2.0] S[table-format=1.5] S[table-format=1.0] S[table-format=1.5] S[table-format=1.1]}
   \hline
    {Resolution} & {$\delta v$} & {n} & {$T_n$}& {Scale Factor}\\ %use {} around text in a column with "S" alignment
    {(mHz)} & {(m/s)} & & {($\sigma_n$)} &\\
   \hline
   \hline
    84 & 15 & 2 & 15.5 & 1.00\\
    168 & 30 & 4 & 18.3 & 1.18\\
    546 & 96 & 13 & 25.4 & 1.64\\
    1092 & 192 & 26 & 31.3 & 2.02\\
   \hline
  \end{tabular*}
 \end{center}
\end{table}

\section{\label{sec:dis}Discussion}

Based on the data recorded between March 31, 2009 and April 1, 2010, the local density of non-virialized, axionic dark matter has been constrained for both KSVZ and DFSZ models at 90$\%$ confidence over the frequency ranges of 812--852 and 858--892 MHz. Respectively, these frequencies correspond to mass ranges of 3.36--3.52 and 3.55--3.69 $\mu$eV. For each resolution, all densities above the limits shown in Fig.~\ref{fig:limits} have been excluded. Equation~\ref{eq:disp} is used with $f = 850$ MHz and $v=300$ km/s to calculate the maximum velocity dispersion for which each resolution is sensitive. These dispersions are included in Table~\ref{tbl:density}. When going from finer to coarser resolutions, the limits become applicable to a larger assortment of axion populations at the cost of becoming less stringent.

Despite this expanded applicability, the tidal flow from the Sagittarius dwarf galaxy is not expected to be seen in even the coarsest resolution, $n=26$. With a velocity dispersion of $\delta v \simeq 20$ km/s \cite{Freese.04, Freese.05}, the expected signal width would be $\Delta f \simeq$ 113 Hz which is more in line with the 2-bin search of the MR channel. Even though the Sagittarius tidal flow is beyond the scope of this analysis, other tidal flows may still be considered using the $n = 26$ data. This resolution would be sensitive to tidal flows with $\delta v \lesssim 192$  m/s. For KSVZ axions, such flows are excluded for $\rho \gtrsim$ 0.8 GeV/cm$^3$ from 812 to 822 MHz, $\rho \gtrsim$ 0.4 GeV/cm$^3$ from 822 to 852 MHz, and $\rho \gtrsim$ 0.2 GeV/cm$^3$ from 859 to 892 MHz.

The disparity between the best and worst limits is largely due to the low values of $Q$ caused by the poor electrical contact between the walls and lid of the cavity during the 2009 data taking. The spectra measured near both 835 MHz and 850 MHz as well as those measured above 860 MHz were all from the 2010 data. Despite the factor of 2 increase in the system noise from bypassing the MSA, the greatly improved $Q$ still yields limits which are a factor of 2 better than the best limits set from the 2009 data. In these regions for $\delta v\lesssim 15$ m/s, KSVZ and DFSZ axions are excluded for densities exceeding $\rho \simeq 0.1$ GeV/cm$^3$ and $\rho \simeq 0.7$ GeV/cm$^3$ respectively.

The local densities of relatively new flows of late-infall axions are each expected to be $\rho \sim 0.01$ GeV/cm$^3$ \cite{Sikivie.97} which is an order of magnitude lower than the best KSVZ limits produced from this data set. Consequently, numerous additional measurements would be required to sufficiently lower the limits so as to exclude late-infall flows of KSVZ axions. Exclusion of flows of DFSZ axions would require many more still.

Finally, proximity to the caustic ring formed by the Big Flow leads to an estimated contribution to the local dark matter density of $\rho \approx 0.9$ GeV/cm$^3$. With a velocity dispersion of $\delta v \simeq 50$ m/s \cite{Sikivie.03}, the Big Flow is most closely matched to the $n = 13$ spectra. For KSVZ axions, the weakest limits produced by the $n = 13$ spectra is 0.8 GeV/cm$^3$. Even in the range of 852--858 MHz, where the present analysis is supplemented by earlier results, the density limits are well below 0.9 GeV/cm$^3$. This permits the exclusion of the Big Flow for KSVZ axions over the entire frequency range from 812 to 892 MHz, equivalent to the mass range of 3.36--3.68~$\mu$eV. For DFSZ axions, the $n = 13$ limits fail to dip below 0.9 GeV/cm$^3$ over any appreciable range. As such, Big Flow can not be excluded for DFSZ axions at this time.

\subsection*{Acknowledgments}
This work was supported by the U.S. Department of Energy through grant numbers DE-SC0010280, DE-FG02-97ER41029, DE-FG02-96ER40956, DE-AC52-07NA27344, and DE-AC03-76SF00098. Additional support was provided by the Heising-Simons Foundation and the Lawrence Livermore National Laboratory LDRD program.

%-------------------------------------------------------------------------------------------------------%

%\bibliographystyle{plain}
%\bibliographystyle{abbrvnat}
%\bibliographystyle{plainnat}
%\bibliographystyle{unsrtnat}
%\bibliographystyle{Chicago_Web}
%\bibliographystyle{apa-good}
%\bibliographystyle{Science_Web}
%\bibliographystyle{ecology_web}
%\bibliographystyle{mla-good}
%\bibliographystyle{mla_web}
%\bibliographystyle{h-physrev} % h-physrev.bst
%\bibliographystyle{apsrev4-1} % apsrev4-1.bst

% \bibliography{refs}

%

\end{document}